\documentclass[showpacs]{revtex4}
\usepackage{graphicx}
\usepackage{amsmath}
\usepackage{mathrsfs}
\usepackage{amsfonts}
\usepackage{amssymb}
\usepackage{bm}
\usepackage{braket}

\newcommand{\fd}{\mathcal{D}}
\newcommand{\ud}{\mathrm{d}}

\newcommand{\Tr}{\mathrm{Tr}}
\newcommand{\dslash}{\partial \!\!\! /}
\newcommand{\sig}{\sigma_0}
\newcommand{\et}{\eta_0}
\newcommand{\condsig}{\braket{\bar \psi \psi}}
\newcommand{\condeta}{\braket{\bar \psi i \gamma_5 \psi}}
\newcommand{\half}{\frac{1}{2}}

\begin{document}

\title{The high temperature CP-restoring phase transition at $\theta = \pi$}

\author{Jorn K. Boomsma}
\author{Dani\"el Boer}
\affiliation{Department of Physics and Astronomy, Vrije Universiteit 
Amsterdam\\
De Boelelaan 1081, NL-1081 HV Amsterdam, the Netherlands}

\date{\today}

\begin{abstract}
  The CP-restoring phase transition at $\theta = \pi$ and high temperature is
  investigated using two related models that aim to describe the low-energy
  phenomenology of QCD, the NJL model and the linear sigma model coupled to
  quarks. Despite many similarities between the models, 
  different predictions for the order of the phase transition result. Using
  the Landau-Ginzburg formalism, the origin of this difference is traced back
  to a non-analytic vacuum term at zero temperature that is present in the NJL
  model, but usually not included in the linear sigma model. Due to the
  absence of explicit CP violation, this term always alters the qualitative 
  aspects of the high temperature phase transition at $\theta = \pi$, just as
  for $\theta=0$ in the chiral limit.
\end{abstract}


\maketitle
It is well known that there is a possibility of CP violation in the
strong interaction due to instanton contributions. These contributions
are incorporated in the QCD Lagrangian through the topological
$\frac{\theta g^2}{32 \pi^2} F \tilde F$-term, where $\theta$ is the
QCD vacuum angle. This term violates CP, unless $\theta = 0~\rm{mod}\ \pi$. 
The case $\theta = \pi$ is special, because then
Dashen's phenomenon can occur, i.e., spontaneous
CP violation at $\theta = \pi$~\cite{Dashen:1970et}.

From experiments it is known that in nature $\theta$ is very
small~\cite{Smith:1990ke,Altarev:1992cf,Jacobs,Vicari:2008jw}. 
The reason for this is unknown and is commonly
referred to as the strong CP problem. However, it has been argued
that in heavy-ion collisions meta-stable CP-violating states could be
created corresponding to
states with an effective $\theta$~\cite{Lee:1973iz,Morley:1983wr,
  Kharzeev:1998kz,Buckley:1999mv,Kharzeev:1999cz,Voloshin:2004vk,
  Kharzeev:2007tn,Kharzeev:2007jp}. Studying the behavior of the
strong interactions at nonzero $\theta$ is therefore of interest and has been  
done quite extensively using chiral Lagrangians, see for
example Refs~\cite{Witten:1980sp,Di
  Vecchia:1980ve,Kawarabayashi:1980uh,
  Smilga:1998dh,Tytgat:1999yx,Creutz:2003xu,Metlitski:2005db}.

Recently the $\theta$-dependence of two models describing the chiral dynamics
of low energy QCD have been studied, the NJL model~\cite{Boer:2008ct} and the
linear sigma model coupled to quarks (LSM$q$)~\cite{Mizher:2008hf}. In both
models the effects of instantons are included through an additional 
interaction, the
't Hooft determinant interaction~\cite{'tHooft:1976fv,'tHooft:1986nc}. It was
found that both models exhibit Dashen's phenomenon, which turns out to be
temperature dependent. This is to be expected, since at high temperature the
effects of instantons, which are needed for the CP violation, are
exponentially suppressed~\cite{Gross:1980br}. In both models the spontaneous
CP violation at $\theta = \pi$ disappears at a critical temperature between 100
and 200 MeV, however, the order of the phase transition differs. In case
of the NJL model the transition is of second order, whereas in the 
LSM$q$ model it is of first order. Clearly this difference is important,
because a first order transition allows meta-stable phases, in contrast to 
a second order transition.

Although the NJL and LSM$q$ model are not the same, they are closely related.
Eguchi~\cite{Eguchi:1976iz} has shown that when the NJL model is
bosonized, a linear sigma model is obtained (see also
\cite{Klevansky:1992qe}). However, the effects of quarks are
treated differently in both models, which was already discussed in
Ref.~\cite{Scavenius:2000qd} for $\theta = 0$. In the case of
the LSM$q$ model the effects of the quarks are usually only taken into
account for nonzero temperatures, whereas in the NJL model their effects are
necessarily incorporated also at zero temperature. 
Ref.~\cite{Scavenius:2000qd}
found that the order of the chiral symmetry restoring phase transition at
$\theta = 0$ was the same in both models, but the critical temperatures
differ. While the qualitative aspects of
the phase transition are similar at $\theta = 0$, this is not the case
for the high temperature CP-restoring phase transition at
$\theta = \pi$ as we will discuss in detail. We should mention here that 
the situation at $\theta = 0$ depends on the amount of explicit chiral symmetry
breaking. In Ref.~\cite{Schaefer:2006ds} it was observed 
that when the pion mass is reduced in order to study the chiral limit, 
neglecting the effects of the
quarks at zero temperature can affect the order of the high temperature phase
transition at $\theta = 0$ too.

Although there is a CP-restoring phase transition at high chemical
potential also, in this paper we will restrict to the temperature dependence of
this phase transition at $\theta = \pi$, because there the differences between
the two models are most pronounced. The paper is organized as follows. 
First, the effective potentials of both models
are analyzed analytically, which will allow the determination of the
order of the phase transitions using standard Landau-Ginzburg type of
arguments. A comparison to numerical results obtained earlier corroborates
these conclusions. Subsequently, we will discuss the bosonification procedure
of Eguchi, which relates the NJL model to a linear sigma model and
allows us to 
further pinpoint the origin of the similarities and differences with the LSM$q$
model.

\section{NJL model}
The Nambu-Jona-Lasinio (NJL) model, introduced in
Refs.~\cite{Nambu:1961tp,Nambu:1961fr}, is a model for low energy QCD 
that contains four-point interactions between the
quarks. In this paper the following form of the NJL model is used,
in the notation of Ref.~\cite{Boer:2008ct}
\begin{equation}
 \mathcal{L}_{\rm NJL} = \bar \psi \left(i \gamma^\mu \partial_\mu - m \right) \psi
   + \mathcal{L}_{\bar q q} + \mathcal{L}_\mathrm{det} \label{lagrangian_NJL},
\end{equation}
where $m$ is the current quark mass. In contrast to
Ref.~\cite{Boer:2008ct}, here the up and down quark masses
are taken equal, which matters little for our present purposes.
Furthermore,
\begin{equation}
 \mathcal{L}_{\bar q q} = G_1 \left[ (\bar \psi \tau_a \psi)^2 +  
 (\bar \psi \tau_a i \gamma_5 \psi)^2 \right],
\end{equation}
is the attractive part of the $\bar q q$ channel of the Fierz transformed
color current-current interaction~\cite{Buballa:2003qv} and
\begin{eqnarray}
 \mathcal{L}_\mathrm{det} & = & 8 G_2 e^{i \theta} 
\det \left( \bar \psi_R \psi_L \right) + \mathrm{h.c.} \label{det_int},
\end{eqnarray}
is the 't Hooft determinant interaction which depends on the QCD
vacuum angle $\theta$ and describes the effects of
instantons~\cite{'tHooft:1976fv,'tHooft:1986nc}. In the literature
$G_1$ and $G_2$ are often taken equal, which at $\theta=0$ means that the
low energy spectrum consists of $\sigma$ and $\bm{\pi}$ fields only, but here 
we will allow them to be different.
We will restrict to the two flavor case, using $\tau_a$ with
$a=0,...,3$ as generators of U(2). We will not consider nonzero
baryon or isospin chemical potential.

The symmetry structure of the NJL model is very similar to that of
QCD. In the absence of quark masses and the instanton interaction,
there is a global SU(3)$_c \times$U(2)$_L \times$U(2)$_R$-symmetry.
The instanton interaction breaks it to SU(3)$_c
\times$SU(2)$_L\times$SU(2)$_R\times$U(1)$_B$. For nonzero, but equal
quark masses this symmetry is reduced to SU(3)$_c
\times$SU(2)$_V\times$U(1)$_B$.

We choose the parameters the same way as in
Refs.~\cite{Boer:2008ct,Frank:2003ve}.  This means we write
\begin{equation}
 G_1 = (1 - c) G_0, \quad G_2 = c G_0, \label{values_G1_G2}
\end{equation}
where the parameter $c$ controls the instanton interaction, while the
value for the quark condensate at $\theta = 0$ (which is determined by
the combination $G_1+G_2$) is kept fixed. For our numerical studies
we will use the following values for the parameters: $m = 6$ MeV, a
three-dimensional momentum UV cut-off $\Lambda = 590$ MeV/$c$ and $G_0
\Lambda^2 = 2.435$. These values lead to a pion mass of $140.2$ MeV, a
pion decay constant of $92.6$ MeV and finally, a quark condensate
$\braket{\bar u u} = \braket{\bar d d} = (-241.5\ \mathrm{MeV})^3$
\cite{Frank:2003ve}, all in reasonable agreement with
experimental determinations.

\subsection{The effective potential}
To calculate the ground state of the theory, the effective potential
has to be minimized. In this section the effective potential is
calculated in the mean-field approximation. In the following we will
only consider the case of unbroken isospin symmetry, such that only 
nonzero $\condsig$ and/or $\condeta$ can arise. 
At $\theta = 0$ only $\condsig$ becomes nonzero. A
nonzero $\condeta$ signals that CP invariance is broken, i.e., 
it serves as an order parameter for the CP-violating phase.

To obtain the effective potential in the mean-field approximation,
first the interaction terms are ``linearized'' in the presence of the
$\condsig$ and $\condeta$ condensates (this is equivalent to the
procedure with a Hubbard-Stratonovich transformation used in
Ref.~\cite{Boer:2008ct})
\begin{eqnarray}
  (\bar \psi \psi)^2 & \simeq & 2 \condsig \bar \psi \psi - \condsig^2, \nonumber \\
  (\bar \psi i \gamma_5 \psi)^2 & \simeq & 2 \condeta \bar \psi i \gamma_5 \psi - \condeta^2, \nonumber \\[.6mm]
  (\bar \psi \psi) (\bar \psi i \gamma_5 \psi) & \simeq & \condsig \bar \psi i \gamma_5 \psi +
  \condeta \bar \psi \psi - \condsig \condeta,
\end{eqnarray}
leading to
\begin{equation}
  \mathcal{L}^{\rm vac}_{\rm NJL} =  \bar \psi \left(i \gamma^\mu \partial_\mu - \mathcal{M} \right) \psi -
    \frac{(G_1 - G_2 \cos \theta) \alpha_0^2}{4(G_1^2 - G_2^2)} -
    \frac{(G_1 + G_2 \cos \theta) \beta_0^2}{4(G_1^2 - G_2^2)} -
    \frac{(G_2 \sin \theta) \alpha_0 \beta_0}{2(G_1^2 - G_2^2)},
\end{equation}
where $\mathcal{M} = (m + \alpha_0) + \beta_0 i \gamma_5$  and
\begin{eqnarray}
  \alpha_0 & = & -2 (G_1 + G_2 \cos \theta) \condsig 
    + 2 G_2 \sin \theta \condeta, \nonumber \\
  \beta_0  & = &  -2 (G_1 - G_2 \cos \theta) \condeta 
    + 2 G_2 \sin \theta \condsig.
\end{eqnarray}
This Lagrangian is quadratic in the quark fields, so the integration
can be performed. After going to imaginary time the thermal effective
potential in the mean-field approximation is obtained~\cite{Warringa:2005jh}
\begin{equation}
  \mathcal{V}^{\rm vac}_{\rm NJL} = \frac{\alpha_0^2 (G_1 - G_2 \cos \theta)}{4 (G_1^2 - G_2^2)} +
  \frac{\beta_0^2 (G_1 + G_2 \cos \theta)}{4 (G_1^2 - G_2^2)} +
  \frac{G_2 \alpha_0 \beta_0 \sin \theta}{2 (G_1^2 - G_2^2)} + \mathcal{V}_q,
\end{equation}
with
\begin{equation}
  \mathcal{V}_q = - T \, N_c \sum_{p_0=(2n+1)\pi T} \int \frac{\ud^3 p}{(2 \pi)^3} \log \det K, \label{effpot}
\end{equation}
and where $K$ is the inverse quark propagator,
\begin{equation}
 K = (i\gamma_0 p_0 + \gamma_i p_i) - \mathcal{M}.
\end{equation}

In order to calculate the effective potential, it is convenient to
multiply $K$ with $\gamma_0$, which does not change the determinant,
but gives a new matrix $\tilde K$ with $i p_0$'s on the diagonal. It
follows that $\det K = \prod_{i=1}^{8} \left(\lambda_i - i p_0
\right)$, where $\lambda_i$ are the eigenvalues of $\tilde{K}$ with
$p_0 = 0$. Because of the symmetries of the inverse propagator, half
of the eigenvalues are equal to $E_{\bm{p}} = \sqrt{\bm{p}^2 + M^2}$
and the other half to $E_{\bm{p}} = - \sqrt{\bm{p}^2 + M^2}$, with
$M^2=(m + \alpha_0)^2 + \beta_0^2$.  After the summation over the
Matsubara frequencies, we obtain
\begin{equation}
  \mathcal{V}_q = - 8 N_c \int \frac{\ud^3 p}{( 2\pi)^3} \left[ \frac{E_{\bm{p}}}{2} + T \log \left(1 + e^{-E_{\bm{p}}/T} \right) \right].
\end{equation}
At $T = 0$ this integral can be performed analytically. A conventional
non-covariant three-dimensional UV cut-off is used to regularize the
integral and yields: 
\begin{equation}
  \mathcal{V}_q^{T=0} = \nu_q \frac{|M| \left(M^3 \log \left(\frac{\Lambda }{M} + \sqrt{1+\frac{\Lambda^2}{M^2}} \right)-\Lambda 
      \left(M^2+2 \Lambda ^2\right) \sqrt{\frac{\Lambda ^2}{M^2}+1}\right)}{32 \pi ^2}, \label{Vq}
\end{equation}
where the degeneracy factor $\nu_q = 24$. 

\subsection{The CP-restoring phase transition}
In this section the high-$T$ CP-restoring phase transition at $\theta = \pi$ is
investigated. As was shown in Ref.~\cite{Boer:2008ct} the
phenomenon of spontaneous CP violation is governed by the strength $c$ of
the 't Hooft determinant interaction. It will be assumed that $c$ is 0.2, 
which following the arguments of Ref.~\cite{Frank:2003ve} is considered 
realistic. But in fact, 
the critical temperature is too very good approximation $c$-independent 
for $c$ above $\sim 0.05$, as can be 
seen from the $(T,c)$ phase diagram given in Ref.~\cite{Boer:2008ct}.

\begin{figure}[htb]
  \includegraphics[scale=0.9]{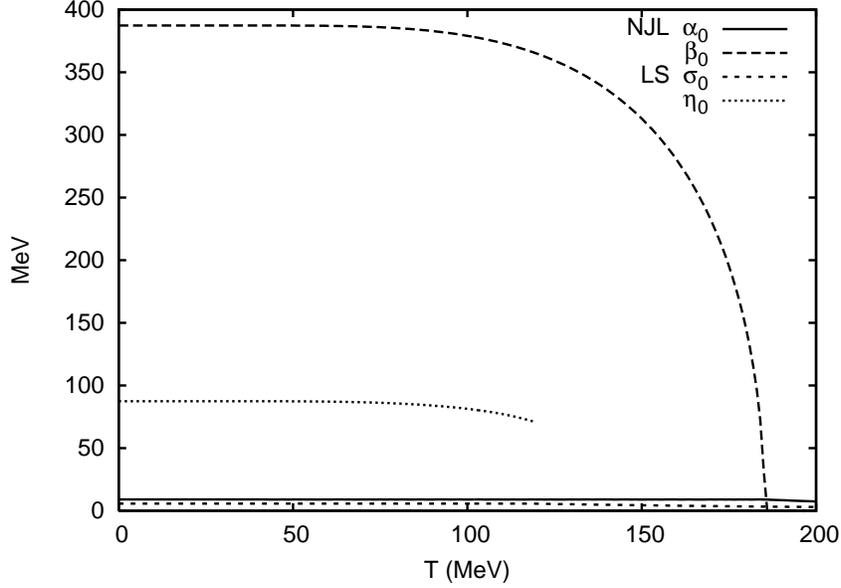}
  \caption{The temperature dependence of the condensates in the NJL
    and linear sigma model.}
  \label{T_dep_cond}
\end{figure}

We will start with a numerical minimization as a function of the
temperature, the results of which, together with those for the LSM$q$ model, 
are shown in Fig.~\ref{T_dep_cond}. One observes that the critical 
temperature of the NJL model is significantly larger than the one of the 
linear sigma model, in agreement with the results of 
Ref.~\cite{Scavenius:2000qd} for the chiral phase transition at $\theta=0$. 
Furthermore, the order of the phase transition is clearly
different, contrary to the results of Ref.~\cite{Scavenius:2000qd} for 
$\theta=0$. 

Next we will derive an analytic expression for the effective potential for 
the NJL model. Two important
observations which can be made from the numerical study will be help. First, 
we note that $\alpha_0$ is very small and constant as long as $\beta_0$ is 
nonzero, which allows us
to approximate $M^2 \approx \beta_0^2$.
Furthermore, $\beta_0$ and hence $M$ can be considered much smaller 
than $\pi T$ and $\Lambda$, allowing expansions. 
These observations simplify our study considerably.

The phase transition occurs for $M$ much smaller than $\Lambda$, so
Eq.~\eqref{Vq} can be expanded in $M/\Lambda$ at $T=0$:
\begin{equation}
 \mathcal{V}_q^{T=0} = \nu_q \left[ -\frac{M^4\log M^2}{64 \pi ^2}+\frac{M^4
\log \left(4 \Lambda^2\right)}{64 \pi ^2}-\frac{M^4}{128 \pi^2} -\frac{\Lambda ^2 M^2}{16 \pi ^2}
-\frac{\Lambda ^4}{16 \pi ^2} + \cdots \right].\label{Vq_expansion}
\end{equation}
For the phase transition, the non-analytic term $M^4 \log M^2$ turns out to
be very relevant. We will
see that it is exactly the absence of this term at finite temperatures in the
NJL model that causes the differences between the two models.

Usually the temperature-dependent part of the potential has to be
evaluated numerically, however when $M < \pi T$ the integral can be
expanded in $M/T$. As can be inferred from Fig.~\ref{T_dep_cond}
it is exactly this regime which is relevant for the phase transition.
Note that the temperature-dependent part of the potential is
UV finite, which means that for this part the cut-off can be
taken to infinity.  In Ref.~\cite{Boer:2008ct} this was not done,
leading to a slightly larger critical temperature. Performing the
expansion, we obtain \cite{Kapusta:2006pm}
\begin{equation}
   \mathcal{V}_q^T = -\nu_q \int \frac{\ud^3 p}{( 2\pi)^3} T \log \left(1 + e^{-E_{\bm{p}}/T} \right) =
   \nu_q \left[ -\frac{7 \pi^2 T^4}{720} + \frac{M^2 T^2}{48} 
    + \frac{M^4}{32 \pi^2} \left( \gamma_E - \frac{3}{4} + \frac{1}{2} 
\log \frac{M^2}{T^2} - \log \pi  \right) + \cdots \right].
\label{VqT_expansion}
\end{equation}
From this expansion one can see that also the temperature dependence
contains a logarithmic term, that will precisely cancel the one of
Eq.~\eqref{Vq_expansion} when added together. 

Using that $M^2 \approx \beta_0^2$, we end up with the effective
potential
\begin{equation}
 \mathcal{V}^{\rm vac}_{\rm NJL}(T) 
= A_{\rm{NJL}}(T) + B_{\rm{NJL}}(T) \beta_0^2 + C_{\rm{NJL}}(T) \beta_0^4,
\label{VNJL}
\end{equation}
where
\begin{eqnarray}
  A_{\rm{NJL}}(T) & = & -\frac{\left(7 \pi ^4 T^4+45 \Lambda ^4\right) \nu_q}{720 \pi ^2}, \\
  B_{\rm{NJL}}(T) & = & \frac{\left(\pi ^2 T^2-3 \Lambda ^2\right) \nu_q}{48 \pi ^2}+\frac{1}{4 G_0}, \\
  C_{\rm{NJL}}(T) & = & \frac{\left(\log \left(4 \Lambda ^2\right)-\log T^2\right) \nu_q}{64 \pi ^2}+\frac{(-1+
   \gamma_E -\log \pi ) \nu_q}{32 \pi ^2}.
\end{eqnarray}
One observes that the logarithm at zero temperature is cancelled by
the logarithm in the temperature dependence. As long as $\beta_0 <
\pi T, \Lambda$ the potential contains no logarithms and is fully
analytic. We note that this expression is the same as   
the chiral limit at $\theta=0$, with $\beta_0$ replaced by $\alpha_0$.  

The phase transition occurs when $B_{\rm{NJL}}(T)$ changes sign. As
the potential is symmetric and quartic in the order parameter, we
conclude (following Landau-Ginzburg arguments) that the phase transition 
is of second order, which the numerical analysis corroborates.
The critical temperature is equal to
\begin{equation}
  T_c^{\rm NJL} = \sqrt{\frac{3 \nu_q G_0 \Lambda^2 - 12 \pi^2}{G_0 \pi^2 \nu_q}} = 185\ \rm{MeV}.
\end{equation}
As long as $T < \frac{2 \Lambda}{\pi} \exp (-1 + \gamma_E) =
246 \, \rm{MeV}$, $C_{\rm NJL}$ is positive, such that higher order terms in
$\beta_0$ are not needed in the analysis.

\section{LSM$q$ model}
The linear sigma model coupled to quarks, like the NJL model, is an effective
low-energy model for QCD
\cite{Pisarski:1996ne,Scavenius:2000qd,Paech:2003fe,Schaefer:2006ds}, 
similar in form to the Gell-Mann-L\'{e}vy model \cite{GellMann:1960np}. 
It is a hybrid model that includes both meson and constituent
quark degrees of freedom, the latter only at nonzero temperature however. 
As was the case in the NJL model, the effects of
instantons are included via the 't Hooft determinant interaction. In this
paper the analysis of Ref.~\cite{Mizher:2008hf} is followed.

We will start with the $T = 0$ case, when only mesons are considered. 
The Lagrangian, which contains
all Lorentz invariant terms allowed by symmetry and renormalizability
has the following form, using a slightly different notation than
Ref.~\cite{Mizher:2008hf}
\begin{eqnarray}
\mathcal{L}_{LS}&=& \frac{1}{2} {\rm Tr}(\partial_\mu\phi^\dagger \partial^\mu \phi)
+ \frac{\mu^2}{2} {\rm Tr}(\phi^\dagger \phi)
- \frac{\lambda_1}{4} [{\rm Tr}(\phi^\dagger \phi)]^2
-\frac{\lambda_2}{4} {\rm Tr}[(\phi^\dagger \phi)^2]
+ \frac{\kappa}{2}[e^{i\theta}\det(\phi) + e^{-i\theta}\det(\phi^\dagger)] \nonumber \\
&&+ \frac{1}{2} {\rm Tr}[\frac{H}{\sqrt{2}}(\phi +\phi^\dagger)] \; ,
\label{LLS}
\end{eqnarray}
where $\phi$ is chiral field, defined as
\begin{equation}
\phi = \frac{1}{\sqrt{2}}(\sigma + i\eta) +
\frac{1}{\sqrt{2}} (\bm{a}_0 + i \bm{\pi}) \cdot \bm{\tau} \; .
\end{equation}
The Lagrangian incorporates both spontaneous and explicit breaking of
chiral symmetry, the latter through the term proportional to $H$. To
study this symmetry breaking, we can concentrate on the potential
corresponding to Eq.\ (\ref{LLS}), expressed in
the meson fields
\begin{eqnarray} \nonumber
\mathcal{V}^{T=0}_{\rm LS}&=& -\frac{\mu^2}{2} (\sigma^2 +\bm{\pi}^2 + \eta^2 + \bm{a}_0^2)  \\ \nonumber
&&-\frac{\kappa}{2} \cos\theta ~(\sigma^2  +\bm{\pi}^2 - \eta^2 - \bm{a}_0^2) \\ \nonumber
&&+ \kappa ~\sin\theta ~(\sigma\eta - \bm{\pi} \cdot\bm{a}_0) - H\sigma \\ \nonumber
&&+ \frac{1}{4}(\lambda_1 + \frac{\lambda_2}{2}) (\sigma^2 + \eta^2 + \bm{\pi}^2 + \bm{a}_0^2)^2 \\
&&+ \frac{2\lambda_2}{4}(\sigma\bm{a}_0 + \eta\bm{\pi} + \bm{\pi} \times\bm{a}_0)^2 \; .
\end{eqnarray}
The spontaneous symmetry breaking manifests itself through nonzero $\sigma$ and
$\eta$ condensates and are obtained by minimizing the
potential. We allow for these condensates by shifting the fields
\begin{equation}
  \sigma \to \sigma_0 + s, \quad \eta \to \eta_0 + h,
\end{equation}
where $\sigma_0$ and $\eta_0$ are the values that
minimize the potential and $s$ and $h$ are the fluctuations. 
These $\sigma_0$ and $\eta_0$ are proportional to the
condensates $\alpha_0$ and $\beta_0$ of the NJL model, respectively.

The potential can now be split in two parts, a vacuum part and one that 
depends on the fluctuations, i.e.,
\begin{equation}
  \mathcal{V}^{T=0}_{\rm LS} = \mathcal{V}^{{\rm vac},T=0}_{LS} + \mathcal{V}^{\rm fluc}_{\rm LS}.
\end{equation}
First we concentrate on the vacuum part, which is given by the
following expression: 
\begin{equation}
  \mathcal{V}^{{\rm vac},T=0}_{LS} = \frac{\lambda}{4} (\sig^2 - v_\theta^2)^2 - H\sig
  +\frac{\lambda}{4}(\et^2 - u_\theta^2)^2 + \kappa ~\sin\theta ~\sig\et
  +\frac{\lambda}{2}\sig^2\et^2 - \frac{\lambda}{4}(v_\theta^4 +
  u_\theta^4),
\end{equation}
where we have defined the combination of couplings $\lambda \equiv
\lambda_1 + \lambda_2/2$, and follow the notation of
Ref.~\cite{Mizher:2008hf}:
\begin{equation}
v_\theta^2\equiv \frac{\mu^2+\kappa ~\cos\theta}{\lambda}\quad ; \quad
u_\theta^2\equiv v_\theta^2 - \frac{2 \kappa}{\lambda} \cos\theta \; .
\end{equation}
This part of the potential determines the phase structure and has to
be compared with the NJL expression~\eqref{Vq_expansion}. The main
difference is that this potential is fully analytic and does not
contain any logarithmic terms.

The part of the potential that depends on the fluctuations is used to
determine the parameters $\mu^2$, $\kappa$, $H$, $\lambda_1$ and $\lambda_2$
in Eq.\ (\ref{LLS}). They are obtained by fitting the masses contained in
$\mathcal{V}^{\rm fluc}_{\rm LS}$ and the pion decay constant at $\theta = 0$
such that the model reproduces the low-energy phenomenology of QCD. At $\theta
= 0$ $\mathcal{V}^{\rm fluc}_{\rm LS}$ has the following form
\begin{eqnarray}
  \mathcal{V}^{\rm fluc}_{\rm LS} & =& \half \left[ m_{\bm{\pi}}^2 \bm{\pi}^2 + m_\sigma^2 s^2 + m_\eta^2 \eta^2 +
    m_{\bm{a}_0}^2 \bm{a}_0^2 \right] \nonumber \\
  && + \left( \lambda_1 + \half \lambda_2 \right) \sigma_0 s \left( s^2 + \bm{\pi}^2 + \eta^2 \right)  
  + \left( \lambda_1 + \frac{3}{2} \lambda_2 \right) \sigma_0 s \bm{a}_0^2 
  + \lambda_2 \sigma_0 \eta \bm{\pi} \cdot \bm{a}_0 \nonumber \\
  && + \left( \frac{1}{4} \lambda_1 + \frac{1}{8} \lambda_2\right) \left( s^2 + \bm{\pi}^2 + \eta^2 + \bm{a}_0^2 \right)^2
  + \half \lambda_2 \left[ \left( s \bm{a}_0 + \eta \bm{\pi} \right)^2 + \left( \bm{\pi} \times \bm{a}_0 \right)^2 \right].
\end{eqnarray}
The masses depend on the parameters of the model as follows:
\begin{eqnarray}
  m_{\bm{\pi}}^2 & = & - \mu^2  - \kappa + \half (2 \lambda_1 + \lambda_2) \sigma_0^2, \nonumber \\
  m_{\sigma}^2 & = & - \mu^2 - \kappa + \frac{3}{2} (2 \lambda_1 + \lambda_2) \sigma_0^2, \nonumber \\
  m_{\bm{a}_0}^2 & = & - \mu^2 + \kappa + (\lambda_1 + \frac{3}{2} \lambda_2) \sigma_0^2, \nonumber \\
  m_{\eta}^2 & = & - \mu^2 + \kappa + (\lambda_1 + \half \lambda_2) \sigma_0^2.
\end{eqnarray}
The mass values used are: $m_{\bm{\pi}} = 138 \rm{MeV}$, $m_{\sigma} = 600 \rm{MeV}$, $m_{\bm{a}_0} = 980 \rm{MeV}$, and $m_{\eta} = 574 \rm{MeV}$.

At nonzero $\theta$, $\eta_0$ becomes nonzero, which alters the
mass relations. Furthermore, cross terms like $\sigma \eta$ become
nonzero, signalling that the mass eigenstates are no longer CP
eigenstates, as discussed for the NJL model in
Ref.~\cite{Boer:2008ct}. As a consequence, the $\sigma$-field mixes with the
$\eta$-field and the $\bm{\pi}$-field mixes with the $\bm{a}_0$-field. We
  will not give these expressions explicitly here. 

\subsection{Nonzero temperature}
In the LSM$q$ model the quarks start to contribute at nonzero temperatures. 
In fact, it is assumed that all the temperature dependence comes from the 
quarks. In Ref.\ \cite{Scavenius:2000qd} it is argued that this approach is
more justified for studying high $T$ phenomena than considering only thermal 
fluctuations of the meson fields, because at high $T$ constituent quarks 
become light and mesonic excitations heavy. For the study of the chiral phase
transition at $\theta=0$ this approach yields results that are 
qualitatively similar to those of the NJL model. 

The part of the LSM$q$ Lagrangian that depends on the quark fields is:
\begin{equation}
{\cal L}_q = \bar \psi \left[i \dslash 
  -g \left( \sigma + i \gamma_5 \eta + \bm{a}_0 \cdot \bm{\tau} + i \gamma_5 \bm{\pi} \cdot \bm{\tau} \right)
  \right] \psi.
\end{equation}
The quark thermal fluctuations are
incorporated in the effective potential for the mesonic sector, by means of 
integrating out the quarks to one loop~\cite{Mizher:2008hf}.
The resulting quark contribution to the potential is given by
\begin{equation}
  \mathcal{V}_q^T = - \nu_q \int \frac{\ud^3 p}{( 2\pi)^3} T \log \left(1 + e^{-E_{\bm{p}}/T} \right).
  \label{quark_contr_LS}
\end{equation}
This expression is equal to the temperature dependent part of the NJL
model Eq.\ (\ref{VqT_expansion}), 
with again $E_{\bm{p}} = \sqrt{\bm{p}^2 + M^2}$ and the
constituent quark mass $M$ depends on the vacuum expectation values
of the meson fields in the following way: $M = g \sqrt{ \left(\sigma_0^2
    + \eta_0^2 \right)}$, where $g$ is the Yukawa coupling between the
quarks and the mesons. A reasonable value for the constituent quark
mass at $\theta = 0$ fixes this coupling constant. In Ref.~\cite{Mizher:2008hf}
(and here) $g = 3.3$ is used, which leads to a cross-over for the chiral phase 
transition as a
function of temperature at $\theta = 0$ and to a constituent quark
mass of approximately $1/3$ of the nucleon mass.

\subsection{The phase transition}
With all parameters fixed, we can study the CP-restoring
phase transition at $\theta = \pi$ in the LSM$q$ model. 
This was studied in detail, along
with other values for $\theta$, in Ref.~\cite{Mizher:2008hf}. There
also the effect of a magnetic field was discussed, which we will not
take into account.

We are now going to follow the same procedure as for the NJL model to study 
the details of the phase transition.
Again, we start the discussion with numerical results of the
minimization of the effective potential, this time the results of
Ref.~\cite{Mizher:2008hf}. They are shown in
Fig.~\ref{T_dep_cond}. From this figure, two simplifying assumptions
can be inferred. First, as was the case for the NJL model, in the
neighborhood of the phase transition $M < \pi T$, allowing
Eq.~\eqref{quark_contr_LS} to be expanded in $M/T$ as in Eq.\
(\ref{VqT_expansion}). Second, $\sigma_0$
is much smaller than $\eta_0$ which means that we can neglect the
$\sigma_0$-dependence. This assumption leads to a small error near
$\eta_0 \approx 0$, but as we checked explicitly this is not important 
since the structure of the extrema of the potential is not altered.

Summing the contributions at zero and nonzero temperature gives the
following form for the effective potential
\begin{equation}
  \mathcal{V}^{\rm vac}_{\rm LS}(T) = A_{\rm LS} (T) + B_{\rm LS} (T) \eta_0^2 + C_{\rm LS} (T) \eta_0^4 + D_{\rm LS} \eta_0^4 \log \eta_0^2,
\end{equation}
where
\begin{eqnarray}
  A_{\rm LS} (T) & = & -\frac{7}{720} \pi ^2 T^4 \nu_q, \\
  B_{\rm LS} (T) & = & \frac{1}{48} \left(g^2 T^2 \nu_q-24 (\mu^2+\kappa)\right), \\
  C_{\rm LS} (T) & = & \frac{1}{32} \left(\frac{\nu_q \left(\log \left(\frac{g}{\pi  T}\right)+\gamma_E
   -\frac{3}{4}\right) g^4}{\pi ^2}+8 \lambda \right),\\
  D_{\rm LS} & = & \frac{g^4 \nu_q}{64 \pi ^2}.
\end{eqnarray}
The form of this potential is clearly different from the one of the NJL
model Eq.\ (\ref{VNJL}), 
the difference being the uncanceled logarithmic term. This term proportional to
$D_{\rm LS}$ will always cause the phase transition to be of first
order. As observed for the NJL model, also in this case the
potential is exactly the same as the chiral limit at $\theta=0$, 
with $\eta_0$ replaced by $\sigma_0$. Beyond the chiral limit the
explicit symmetry breaking term $\sim H\sigma_0$ 
(which has no analogue at $\theta=\pi$) 
will change the first order transition into a
cross-over, unless the Yukawa coupling $g$ is increased sufficiently
\cite{Paech:2003fe,Schaefer:2006ds}. We conclude that the absence of explicit
CP violation through a linear term in $\eta_0$ at $\theta=\pi$ lies at the 
heart of the difference between the observations made here and those in Ref.\
\cite{Scavenius:2000qd}.

Like in the NJL model, it is the sign flip of $B_{\rm LS}$ that
modifies the structure of the minima. But instead of a phase
transition, now a meta-stable state develops at $\eta_0 = 0$.  When
$B_{\rm LS}(T)$ becomes larger than $2 D_{\rm LS} \exp(-\frac{3}{2}
- \frac{C_{\rm LS}(T)}{D_{\rm LS}})$ the original minimum
disappears. Between the two spinodals the minimum jumps, signalling a
first order transition.

When the parameters of Ref.~\cite{Mizher:2008hf} are used, we obtain
the following values for the spinodals: 118 MeV and 129 MeV. To find
the exact point of the phase transition, the potential has to be
minimized numerically, giving a critical temperature of 126.4
MeV. As
already noted, this is significantly lower than $T_c^{\rm NJL}$,  
but the specific values depend on the parameter choices made.
As should
be clear from the previous discussion, choosing different parameters would
not affect the conclusion about the different orders of the phase
transition, at least as long as $M < \pi T, \Lambda$ and $\kappa > -\mu^2$
(equivalently, $m_\sigma^2 > 3 m_\pi^2$ at $T=0$). 

\section{Relation between the NJL and LSM$q$ model}
As mentioned, the LSM$q$ model is a hybrid model for mesons, which are coupled to
quarks at nonzero temperature, and the
NJL model is a quark model, where the bosonic states of
quark-antiquark fields are interpreted as mesons.
Eguchi~\cite{Eguchi:1976iz} has shown how to derive from the
Lagrangian of the NJL model a Lagrangian for the mesonic excitations
for $G_2 = 0$. This bosonification procedure is reviewed in
Ref.~\cite{Klevansky:1992qe}. Here the corresponding meson Lagrangian
will be derived for $G_2 \neq 0$, which was also studied in
Ref.~\cite{Dmitrasinovic:1996fi} in the chiral limit.

The situation will be reviewed for $\theta=0$, when only the $\bar \psi
  \psi$ receives a vacuum expectation. We start
with the Lagrangian given in Eq.~\eqref{lagrangian_NJL}.  The
generating functional is given by the standard expression
\begin{equation}
  Z[ \bar \xi, \xi] = \frac{1}{N} \int \fd \psi \fd \bar \psi \exp \left(i \int \ud^4 x 
      \left[ \mathcal{L_{\rm NJL}}(\bar \psi, \psi) + \bar \psi \xi + \bar \xi \psi \right] \right),
\end{equation}
where $\bar \xi$ and $\xi$ are the antifermion and fermion sources and
$N$ is a normalization factor which will be suppressed from now on.
Next we introduce auxiliary fields $\sigma$, $\eta$, $\bm{\pi}$ and
$\bm{a}_0$ and a new Lagrangian $\mathcal{L'}$ such that the effective
potential can be written as
\begin{equation}
  Z[ \bar \xi, \xi] = \int \fd \psi \fd \bar \psi \fd \sigma \fd \eta \fd \bm{\pi} \fd \bm{a}_0 \exp \left(i \int \ud^4 x 
    \left[ \mathcal{L}'_{\rm NJL}(\bar \psi, \psi) + \bar \psi \xi + \bar \xi \psi \right] \right),
\end{equation}
with 
\begin{equation}
  \mathcal{L}'_{\rm NJL} = \bar \psi \left[i \dslash - m
  -g \left( \sigma + i \gamma_5 \eta + \bm{a}_0 \cdot \bm{\tau} + i \gamma_5 \bm{\pi} \cdot \bm{\tau} \right)
  \right] \psi - \half \delta \mu_1^2 \left( \sigma^2 + \bm{\pi}^2
  \right) -  \half \delta \mu_2^2 \left( \eta^2 + \bm{a}_0^2 \right),
\end{equation}
and
\begin{equation}
  \delta \mu_1^2 = \frac{g^2}{2(G_1 + G_2)}, \quad  \delta \mu_2^2 = \frac{g^2}{2(G_1 - G_2)}.
\end{equation}
Here $g$ is again the Yukawa coupling between the quarks and mesons, which
in the case of the NJL model can be evaluated. It is equal to
\begin{equation}
g^{-2} =  - 4 N_c i \int \frac{\ud^4 p}{(2 \pi)^4} \frac{1}{(p^2 - M^2)^2},
\label{gLambda}
\end{equation}
which requires some regularization.

Integrating out the quarks gives the following generating functional
\begin{equation}
  Z[ \bar \xi, \xi] = \int\fd \sigma \fd \eta \fd \bm{\pi} \fd \bm{a}_0 
  \exp \left(i \mathcal{S_{\rm NJL}} + i \int \ud^4 x \bar \xi \frac{1}{i \dslash 
      - m  -g \left( s + i \gamma_5 \eta + \bm{a}_0 \cdot \bm{\tau} + i \gamma_5 \bm{\pi} \cdot \bm{\tau} \right)} \xi \right)
\end{equation}
where the action $\mathcal{S_{\rm NJL}}$ is equal to
\begin{equation}
  \mathcal{S_{\rm NJL}} = \int \ud^4 x \left[  - \half \delta \mu_1^2 \left( \sigma^2 + \bm{\pi}^2
    \right) -  \half \delta \mu_2^2 \left( \eta^2 + \bm{a}_0^2 \right) \right] 
  - i \Tr \log \left[ i \dslash - m
    -g \left( \sigma + i \gamma_5 \eta + \bm{a}_0 \cdot \bm{\tau} + i \gamma_5 \bm{\pi} \cdot \bm{\tau} \right)
  \right].
\end{equation}
Assuming that only the $\sigma$-field receives a vacuum expectation
value $\sigma_0$, i.e., $\sigma=\sigma_0+s$, the action can be split 
into a vacuum part and a part that depends on
the fluctuations, which are the mesons $s, \eta, \bm{\pi}, \bm{a}_0$: 
\begin{equation}
  \mathcal{S_{\rm NJL}} = \mathcal{S}^{\rm vac}_{\rm NJL} + \mathcal{S}^{\rm fluc}_{\rm NJL},
\end{equation}
with
\begin{eqnarray}
  \mathcal{S}^{\rm vac}_{\rm NJL} &  = & \int \ud^4 x \left[  - \half \delta \mu_1^2 \sigma_0^2 \right]
    - i \Tr \log \left[ i \dslash - M \right], \nonumber \\
  \mathcal{S}^{\rm fluc}_{\rm NJL} & = & \int \ud^4 x \left[  
      - \half \delta \mu_1^2 \left( s^2 + 2 \sigma_0 s  + \bm{\pi}^2 
      \right) -  \half \delta \mu_2^2 \left( \eta^2 + \bm{a}_0^2 \right) \right] \nonumber \\ 
   && - i \Tr \log \left[ 1 - \frac{1}{i \dslash - M}
      g \left( s + i \gamma_5 \eta + \bm{a}_0 \cdot \bm{\tau} + i \gamma_5 \bm{\pi} \cdot \bm{\tau} \right)
    \right],
\end{eqnarray}
and the constituent quark mass $M=m + g \sigma_0$. In order to obtain
  a local action for the meson fields, the nonlocal fermionic determinant
in $\mathcal{S}^{\rm fluc}_{\rm NJL}$ is rewritten using a  
derivative expansion:
\begin{equation}
   - i \Tr \log \left[ 1 - \frac{1}{i \dslash - M}
      g \left( s + i \gamma_5 \eta + \bm{a}_0 \cdot \bm{\tau} + i \gamma_5 \bm{\pi} \cdot \bm{\tau} \right)
    \right] = \sum_{n=1}^\infty U^{(n)},
\end{equation}
where
\begin{equation}
  U^{(n)} = \frac{1}{n} \Tr \left( \frac{1}{i \dslash - M} g \left(s + i \gamma_5 \eta + \bm{a}_0 \cdot \bm{\tau} + i \gamma_5 \bm{\pi} \cdot \bm{\tau} \right) \right)^n.
\end{equation}
From power counting we note that $U^{(n)}$ with $n \geq 5$ are
convergent and the rest is divergent. Evaluating and retaining only 
the divergent parts
of the $U^{(n)}$ with $n = 1,2,3,4$ we end up with the following
Lagrangian, which integrated over all space yields 
$\mathcal{S}^{\rm fluc}_{\rm NJL}$:
\begin{eqnarray}
  \mathcal{L}^{\rm fluc}_{\rm NJL} & =& \half \left[ (\partial_\mu s)^2 + (\partial_\mu \eta)^2 + (\partial_\mu \bm{a}_0^2)^2 +
      (\partial_\mu \bm{\pi})^2 \right] - \half \left[ m_{\bm{\pi}}^2 \bm{\pi}^2 + m_\sigma^2 s^2 + m_\eta^2 \eta^2 +
      m_{\bm{a}_0}^2 \bm{a}_0^2 \right] \nonumber \\
    && - g_3 s \left( s^2 + \bm{\pi}^2 + \eta^2 + 3 \bm{a}_0^2 \right) - 2 g_3 \eta \bm{\pi} \cdot \bm{a}_0
    - \half g_4 \left( s^2 + \bm{\pi}^2 + \eta^2 + \bm{a}_0^2 \right)^2 \nonumber \\
    && - 2 g_4 \left[ \left( s \bm{a}_0 + \eta \bm{\pi} \right)^2 + \left( \bm{\pi} \times \bm{a}_0 \right)^2 \right].
    \label{L_mesons_NJL}
\end{eqnarray}
The masses and coupling constants have the following values
\begin{eqnarray}
   m_{\bm{\pi}}^2 & = & \frac{1}{2 G_0 I_0} - 2 \frac{I_2}{I_0} = \frac{m}{M} \frac{1}{2 G_0 I_0},  \nonumber \\
   m_\sigma^2 & = & m_{\bm{\pi}}^2 + 4 M^2, \nonumber \\
   m_\eta^2 & = & \frac{1}{2 (1 - 2c) G_0 I_0} - 2 \frac{I_2}{I_0}, \nonumber \\
   m_{\bm{a}_0}^2 & = & m_\eta^2 + 4 M^2, \nonumber \\
   g_3 & = &\frac{2 M}{I_0^{1/2}} = 2 M g_4^{1/2}, \nonumber \\
   g_4 & = & \frac{1}{I_0}, \nonumber \\
   g & = & \frac{1}{I^{1/2}_0}, \label{masses_LS_from_NJL}
\end{eqnarray}
where $I_0$ and $I_2$ are two divergent integrals, equal to
\begin{eqnarray}
  I_0 & = & - 4 N_c i \int \frac{\ud^4 p}{(2 \pi)^4} \frac{1}{(p^2 - M^2)^2}, \nonumber \\
  I_2 & = & 4 N_c i \int \frac{\ud^4 p}{(2 \pi)^4} \frac{1}{p^2 - M^2}.
\end{eqnarray}
If these integrals are regularized using the three-dimensional UV
cut-off, the resulting masses are equal to the ones obtained using the
random phase approximation used in Ref.~\cite{Boer:2008ct} (where
  the dependence on the external momentum of the generalized $I_0$
  defined in, for example, Ref.~\cite{Klevansky:1992qe} has been
  neglected). The Lagrangian~(\ref{L_mesons_NJL}) without the
$\bm{a}_0$ and $\eta$-fields was also given in
Ref.~\cite{Ebert:1982pk}. In the chiral limit the results agree with
those of Ref.~\cite{Dmitrasinovic:1996fi}.

Eq.~\eqref{L_mesons_NJL} is equal to the fluctuation part of the
linear sigma model Lagrangian (\ref{LLS}) using the following parameters
\begin{eqnarray}
  \lambda_1 & = & 0, \nonumber \\
  \lambda_2 & = & 4 / I_0, \nonumber \\
  \mu^2 & = & 2 M^2 - \frac{c}{2 (1-2c)G_0 I_0}, \nonumber \\
  \kappa & = & \frac{c}{2 (1-2c)G_0 I_0}, \nonumber \\
  H & = & \frac{m}{2 G_0 I_0^{1/2}}.
\end{eqnarray}
Although the bosonification of the NJL model yields $\lambda_1 = 0$,
this is of no consequence for the order of the phase transition, as
the effective potential at zero temperature is a quartic polynomial
irrespective of whether $\lambda_1 = 0$. It does however, affect the
masses of the mesons. If $\lambda_1 = 0$, the following relation
holds: $m_\sigma^2 - m_{\bm{\pi}}^2 = m_{\bm{a}_0}^2 - m_\eta^2 = 4
M^2$, a property of the NJL model already noted in
Ref.~\cite{Dmitrasinovic:1996fi}.  Clearly, the bosonized NJL model
does not yield the most general linear sigma model.  However, it gives
additional contributions to the vacuum that usually are not taken into
account in the linear sigma model coupled to quarks
\cite{Scavenius:2000qd,Mizher:2008hf}. In Ref.~\cite{Schaefer:2006ds}
it is noted that upon inclusion of fluctuations using an RG flow
equation, the transition becomes second order. This boils down to
including quark loop effects at zero temperature too and is consistent
with our findings.

To conclude, the mesonic part of this bosonized NJL Lagrangian is
equal to the mesonic part of the LSM$q$ model. So the mesons are
treated in same way in the two models, but the vacuum contributions are
treated differently. Since neither model is directly derived from QCD,
it is not straightforward to draw a conclusion about the order of the
phase transition expected in QCD. If the NJL model is viewed as a
model for the 
microscopic theory underlying the low energy mesonic theory, it would
not seem justified to neglect the logarithmic term at zero
temperature.

It is straightforward to bosonize the NJL model for $\theta \neq 0$
when 
also $\braket{\psi i \gamma_5 \psi}$ can become nonzero, leading
to cross terms that mix the $\sigma$-field with the $\eta$-field and
$\bm{a}_0$-field with the $\bm{\pi}$-field, but we do not give the
  expressions here as they do not lead to any additional insights.

\section{Conclusions}
In this paper the high-$T$ CP-restoring phase transition at $\theta =
\pi$ was discussed for two different models which aim to describe the
low-energy QCD phenomenology, the NJL model and the linear sigma model
coupled to quarks. Although the models are related, the philosophy of
how the mesons are treated is quite different in both models. In the
NJL model they are bosonic states of quark-antiquarks, whereas in the
LSM$q$ model they are the fundamental degrees of freedom, interacting
with quarks at nonzero temperature. Using the bosonification procedure
of Eguchi, one can show that a bosonized NJL model gives a linear
sigma model, in which mesons are treated in the same way as in the
LSM$q$ model. However, the vacuum contributions arising from the quark
degrees of freedom are different. The LSM$q$ model
was motivated for high temperatures, when constituent quarks
are light and mesons are heavy. Therefore, it is assumed that quarks 
only play a
role at nonzero temperature and do not affect the vacuum contributions
at zero temperature. On the other hand, in the NJL model contributions by
the quarks are necessarily taken into account also at zero temperature. The
temperature dependent contributions to the effective potential are
equal in both models, coming
exclusively from the quarks. 
In the end, the effective potentials of
the models only differ in their zero temperature contributions.
Nevertheless, this directly affects the nature of the phase
transition at high temperature at $\theta = \pi$.

The temperature dependence of the ground state of both models was
investigated using a Landau-Ginzburg analysis. The difference between
the models is that the potential as a function of the order parameter
of the LSM$q$ model contains a non-analytic logarithmic term,
whereas the potential of the NJL model is a quartic polynomial near
the phase transition. It is this logarithm that makes the difference,
it affects the order of the phase transition. This logarithm
comes from the contribution of the quarks at zero temperatures, but
neglecting these contributions will affect the high temperature results
qualitatively at $\theta = \pi$. A similar effect occurs for the chiral
symmetry restoration phase transition at $\theta=0$ close to the chiral limit,
i.e.\ for sufficiently small explicit symmetry breaking. The absence of
explicit CP violation is therefore an important aspect of the physics at 
$\theta=\pi$. 

\begin{acknowledgments}
We thank Eduardo Fraga and Ana Mizher for fruitful discussions. 
\end{acknowledgments}

\end{document}